\pdfoutput=1
\documentclass[usenatbib]{mnras}

\usepackage{txfonts}

\usepackage[T1]{fontenc}
\usepackage{ae,aecompl}

\usepackage{graphicx}	
\usepackage{amssymb}	

\title[Dust temperature vs. environment]{Environmental impacts on dust temperature of star-forming galaxies in the local Universe}

\author[Y. Matsuki et al.]{
Yasuhiro Matsuki$^{1,2}$\thanks{E-mail: matsuki@ir.isas.jaxa.jp},
Yusei Koyama$^{3,4}$,
Takao Nakagawa$^{1}$
and Satoshi Takita$^{1}$
\\
$^{1}$Institute of Space Astronautical Science, Japan Aerospace Exploration Agency, Sagamihara, Kanagawa 252-5210, Japan\\
$^{2}$Department of Physics, Graduate School of Science, The University of Tokyo, Tokyo 113-0033, Japan\\
$^{3}$Subaru Telescope, National Astronomical Observatory of Japan, National Institutes of Natural Sciences, 650 North A'ohoku Place,\\ Hilo, HI 96720, USA \\
$^{4}$Graduate University for Advanced Studies (SOKENDAI), Osawa 2-21-1, Mitaka, Tokyo 181-8588, Japan}

\date{Accepted XXX. Received YYY; in original form ZZZ}

\pubyear{2015}

\begin{document}
\label{firstpage}
\pagerange{\pageref{firstpage}--\pageref{lastpage}}
\maketitle

\begin{abstract}
We present infrared views of the environmental effects on the dust
properties in star-forming (SF) galaxies at $z\sim0$, using the
AKARI Far-Infrared Surveyor (FIS) all-sky map and the large spectroscopic galaxy sample from
Sloan Digital Sky Survey (SDSS) Data Release 7 (DR7). We restrict the
sample to those within the redshift range of $0.05<z<0.07$ and the
stellar mass range of $9.2 < \log_{10}(M_*/M_\odot)$. We select SF
galaxies based on their H$\alpha$ equivalent width ($EW_{\rm H\alpha}>$ 4 \AA)
 and emission line flux ratios. We perform far-infrared (FIR) stacking
analyses by splitting the SDSS SF galaxy sample according to their
stellar mass, specific $SFR$ ($SSFR_{\rm SDSS}$), and environment. We
derive total infrared luminosity ($L_{\rm IR}$) for each subsample
using the average flux densities at WIDE-S (90 $\mu$m) and WIDE-L (140
$\mu$m) bands, and then compute IR-based $SFR$ ($SFR_{\rm IR}$) from
$L_{\rm IR}$. We find a mild decrease of IR-based $SSFR$
($SSFR_{\rm IR}$) amongst SF galaxies with increasing local density
($\sim0.1$-dex level at maximum), which suggests that environmental
effects do not instantly shut down the SF activity in galaxies. We
also derive average dust temperature ($T_{\rm dust}$) using the
flux densities at 90 $\mu$m and 140 $\mu$m bands. 
We confirm a strong positive correlation between $T_{\rm dust}$ and 
$SSFR_{\rm IR}$, consistent with recent studies.  
The most important finding of this study is that we find a marginal 
trend that $T_{\rm dust}$ increases with increasing
environmental galaxy density. Although the environmental trend is much milder
than the $SSFR$--$T_{\rm dust}$ correlation, our results suggest 
that the environmental density may 
affect the
dust temperature in SF galaxies, and that the physical mechanism which is
responsible for this phenomenon is not necessarily specific to cluster
environments because the environmental dependence of $T_{\rm dust}$
holds down to relatively low-density environments.
\end{abstract}

\begin{keywords}
galaxies: star formation -- ISM: dust.
\end{keywords}



\section{Introduction}

It is widely known that various galaxy properties, such as star formation 
rate ($SFR$) or morphologies, strongly depend on the local galaxy 
environment \citep{2009ARA&A..47..159B}. 
In the local Universe, \citet{1980ApJ...236..351D} investigated 55 rich 
clusters and found that the fraction of elliptical galaxies increases 
with local galaxy density, while that of spiral galaxies decreases 
towards high-density regions.
\citet{2003MNRAS.346..601G} showed similar results based on the Sloan 
Digital Sky Survey (SDSS) Early Data Release.
\citet{1997ApJ...488L..75B} found that the average $SFR$ for cluster 
galaxies is lower than the average for field galaxies.
\citet{2002MNRAS.334..673L} and \citet{2003ApJ...584..210G} also showed 
that $SFR$ decreases with increasing local galaxy density. 

For star-forming (SF) galaxies, it is established that $SFR$ increases with stellar mass ($M_*$).
This $SFR$--$M_*$ relation for SF galaxies is often called the ``SF main sequence'', and recognized 
both in the local Universe \citep{2004MNRAS.351.1151B,2010ApJ...721..193P} and in the distant 
Universe \citep{2007ApJ...670..156D,2007ApJ...660L..43N,2012ApJ...754L..29W}.
\citet{2010ApJ...721..193P} studied local SF galaxies using SDSS seventh Data Release 
(DR7) and found that the $SFR$--$M_*$ relation is independent of the environment.
\citet{2012MNRAS.423.3679W} showed a similar result by using galaxies observed 
in the Galaxy And Mass Assembly (GAMA), and concluded that the $SFR$--density 
relation is largely driven by the higher fraction of passive early-type galaxies 
in high-density environments. The environmental independence of the SF
main sequence is also suggested for high-redshift galaxies out to $z\sim 2$ 
\citep{2013MNRAS.434..423K}. These results claiming small 
(or lack of) environmental dependence of the SF main sequence may suggest 
a rapid SF quenching mechanism at work in high-density environments. 

However, environmental variations in the properties of star-forming galaxies
is still under debate. 
Some recent studies showed small, but significant 
environmental dependence of SF activity amongst SF galaxies.
\citet{2013ApJ...775..126H} found that specific $SFR$ ($SSFR$) of SF cluster 
galaxies is lower than that of SF galaxies in field environments by using a 
sample of 30 massive galaxy clusters at $0.15<z<0.30$ from the Local Cluster 
Substructure Survey (LoCuSS). Similarly, by using the data from Herschel 
Astrophysical Terahertz Large Area Survey (H-ATLAS), \citet{2016MNRAS.tmp...95F} 
showed that $SSFR$ of late-type galaxies in the Coma cluster declines with 
increasing their local galaxy number density. 
In $0.6<z<0.8$, \citet{2010ApJ...710L...1V} found the $SSFR$ reduction of star-forming galaxies of the same mass in cluster environments by using 24 $\mu$m MIPS/Spitzer data.
\citet{2016MNRAS.tmp...95F} 
also found that gas-to-stars ratio decreases with increasing environmental density.
These results can be explained if the gas content of galaxies is stripped 
when they enter cluster environments.

The gas removal (or reduction) of galaxies is one of the main causes 
for SF quenching in galaxies. Although the dominant mechanism responsible for quenching 
is still under debate, many possible physical mechanisms are proposed as the 
driver of environmental effects: e.g.\ ram-pressure stripping of the cold gas due to 
interaction with the intracluster medium \citep{1972ApJ...176....1G,2000Sci...288.1617Q}, 
galaxy harassment through high velocity encounters with other galaxies 
\citep{1999MNRAS.304..465M}, suppression of the accretion of cold gas: 
$strangulation$ \citep{2011ApJ...736...88F,2009MNRAS.399.2221B,2008ApJ...672L.103K,1980ApJ...237..692L}, 
and galaxy mergers or close tidal encounters of galaxies in in-falling groups 
\citep{1998ApJ...496...39Z}. 

Understanding the dust properties in galaxies is very important 
when we discuss SF activity in galaxies because 
dust absorbs a large fraction of the UV light emitted by O/B-type stars 
and reradiates it in far-infrared (FIR). The H$\alpha$ emission, which is 
often used as a good indicator of $SFR$, can also be significantly attenuated 
by interstellar dust grains in the case of extremely dusty galaxies 
\citep{2000ApJ...529..157P,2010MNRAS.403.1611K}.
FIR observations are thus crucial to study SF activity hidden 
by dust. In particular, the dust temperature ($T_{\rm dust}$) 
can be an important factor because $T_{\rm dust}$ 
is expected to be linked to the physical conditions prevailing in the 
SF regions within galaxies. In fact, it is suggested that galaxies 
with higher $SSFR$ tend to have higher dust temperature due to the 
strong ultraviolet (UV) radiation fields of the young massive stars 
\citep{2014A&A...561A..86M}.

It is shown that the distribution of dust component within galaxies 
well traces that of molecular gas contents \citep{2012A&A...540A..52C}.
Also, dust temperatures are typically higher in the central part of galaxies 
than in the outskirts \citep{2010A&A...518L..56E}.
It is expected that stripping effects tend to remove gas and dust 
from outskirts of galaxies \citep{2007ApJ...667..859D}.
If molecular gas of galaxies is really stripped in high-density 
regions, cold dust in the outskirts of SF galaxies could also be 
stripped, which could result in warm dusts in the inner parts of 
galaxies being exposed. Therefore, we expect that dust temperatures 
of SF galaxies increase with their local densities. 

A few recent studies actually claim a correlation between $T_{\rm dust}$ 
and environments. \citet{2012ApJ...756..106R} studied $z\sim 0.3$ galaxy clusters with 
{\it Herschel}, and found that {\it warm dust} galaxies are preferentially 
located in cluster environments. They attribute this result to $cold$ 
dust stripping effects at work in cluster environments. 
In contrast, \citet{2015arXiv151100584N} showed that $T_{\rm dust}$ tends
to be lower in the cluster environments at $z\sim 1$. They interpreted 
this as a result of $warm$ dust stripping effects. There is no 
systematic study on the environmental effects on dust temperatures
in galaxies so far. This is primarily because dust temperature measurements 
require multi-band photometry at FIR, and that huge area survey at FIR 
is required if we really want to perform studies on environmental 
effects on dust temperatures of galaxies in an unbiased way.

In this paper, we study environmental effects on {\it hidden} SF 
activities in galaxies, particularly focusing on IR-based $SSFR$ and 
dust temperatures, by using the newly released AKARI FIR
all-sky survey map and SDSS DR7 spectroscopic sample. 
Because of the limited depths of FIR all-sky survey performed by AKARI, 
we will focus on the average properties of galaxies by exploiting FIR 
stacking analyses. 

This paper is organized as follows. 
In Section~\ref{data}, we describe the outline of our sample selection, 
local density measurements, as well as the technique of our stacking analyses. 
In Section~\ref{MS}, we present the dependence of $SSFR$ on local density.
The main result of this study is presented in Section~\ref{Td}, where we 
study the environmental dependence of average dust temperature of galaxies
in the local Universe. Because it is expected that dust temperature is 
strongly correlated with $SSFR$ \citep{2014A&A...561A..86M}, we also 
examine the environmental dependence of $T_{\rm dust}$ at fixed $SSFR$.
We compare our results with some recent studies and discuss possible 
explanations for the dependence of $T_{\rm dust}$ on local density.
In Section~\ref{summary}, we summarize our conclusions.

Throughout the paper, we adopt $\Omega_{\rm M}=0.3,\ \Omega_\Lambda=0.7$ and 
$H_0=70\ {\rm km\,s^{-1}\,Mpc^{-1}}$. We assume \citet{2001MNRAS.322..231K} 
initial mass function (IMF) to derive physical quantities in order to keep 
consistency with those derived in the literatures.

\section{data}\label{data}
\subsection{Sloan Digital Sky Survey}
\subsubsection{Definition of environment}\label{def_environment}
In this work, we define the galaxy environment by computing the local density of galaxies. We here describe the outline of our density measurements. 

We use the SDSS (DR7; \citeauthor{2009ApJS..182..543A}\,\,\citeyear{2009ApJS..182..543A}) spectroscopic data. Some galaxy properties (e.g. redshift or stellar mass) are derived by Max Planck Institute for Astropysics and Johns Hopkins University group (MPA/JHU group). We retrieve their ``value-added''  catalogue (hereafter MPA/JHU catalogue) from their website\footnote{http://wwwmpa.mpa-garching.mpg.de/SDSS/DR7}. The catalogue contains a magnitude limited sample of 927,552 galaxies (with \citealt{1976ApJ...209L...1P} magnitude limit of $r' \leq17.77$).

SDSS covers a contiguous area of $\sim 7500\ {\rm deg}^2$ in Northern Galactic Cap and the three stripes in the Southern Galactic Cap. We select a sample of 771,693 galaxies within the area of $105^\circ<\rm{R.A.}<270^\circ$ and $-5^\circ<\rm{DEC.}<75^\circ$, corresponding to the contiguous region, which is better suited for environmental studies. Then, we exclude duplicated objects by performing internal matching with maximum separation of $1''$, yielding 738,143 objects as unique sources. 

We calculate the local density of each galaxy using projected $n$-th nearest neighbour surface density $\Sigma_n$, which is expressed as
\begin{eqnarray}
\Sigma_n=\frac{n}{\pi {D_{p,n}}^2}\ \ ({\rm Mpc^{-2}}),
\end{eqnarray}
where $D_{p,n}$ is the projected comoving distance to the $n$-th nearest neighbour within a velocity window of $\pm$1000~$\rm{km\,s^{-1}}$, or equivalently a redshift slice of $\Delta z=\pm 0.003$. Note that the size of this velocity window ($\pm$1000~$\rm{km\,s^{-1}}$) corresponds to a typical velocity dispersion of galaxy clusters, and so we believe that it is wide enough to pick out the physically related galaxies. 

Because the original sample is magnitude limited, we need to take into account the redshift dependence of the completeness limits. We therefore define the normalized local galaxy number density ($\rho_n$) as follows:
\begin{eqnarray}
\rho_n=\frac{\Sigma_n}{\langle\Sigma_n\rangle},
\end{eqnarray}
where $\langle\Sigma_n\rangle$ is the median $\Sigma_n$ of galaxies at each redshift within a slice of $\Delta z=\pm0.003$. In this paper, we fix to $n=5$.

\subsubsection{Selection of star-forming galaxies}
\begin{figure}
\begin{center}
\includegraphics[width=80mm]{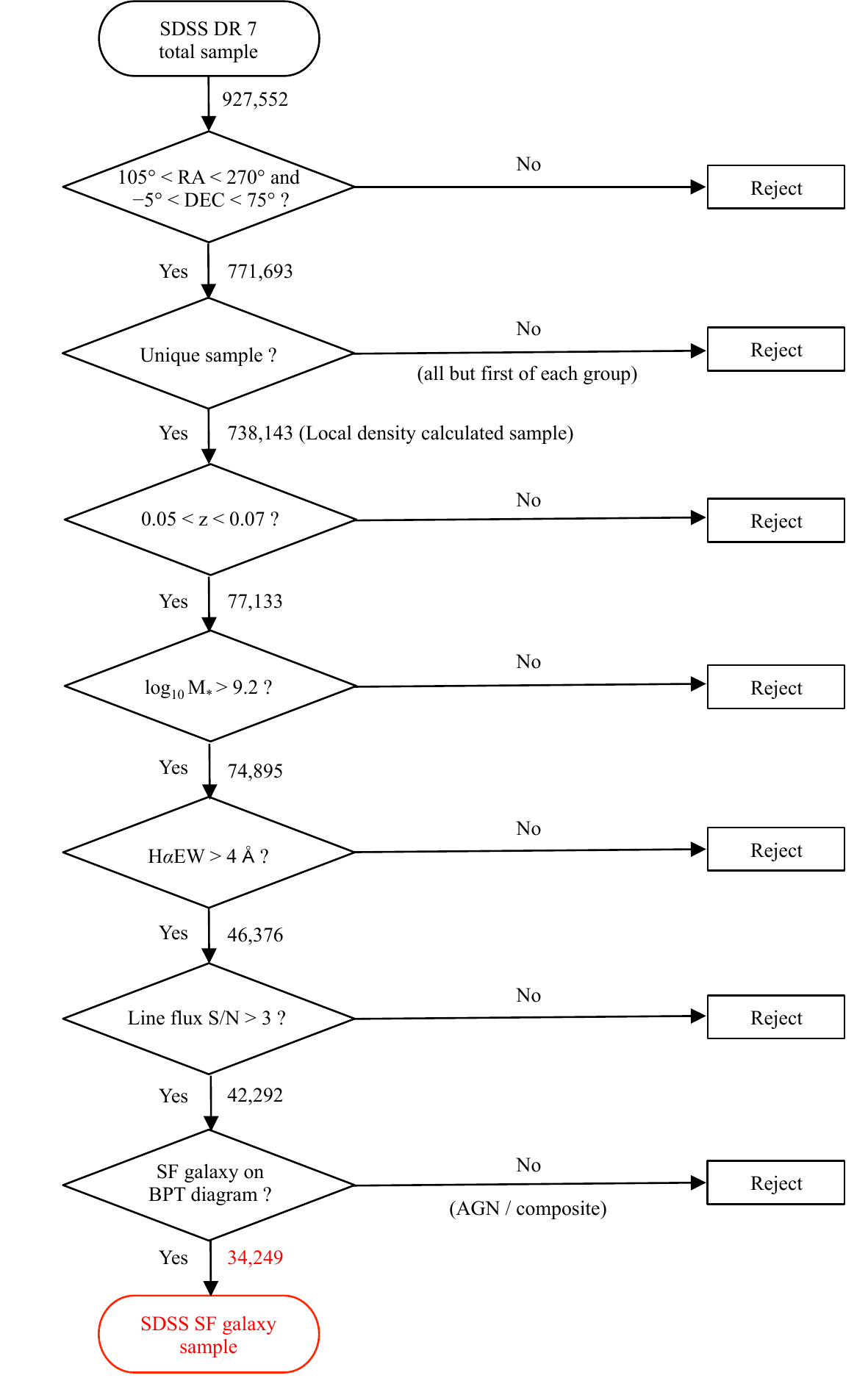}
\end{center}
\caption{Summary of our sample selection procedure.}
\label{sampleSelec}
\end{figure}

In this section, we describe the outline of our sample selection. Our procedure is also summarized in Fig.~\ref{sampleSelec}.

We first restrict the sample to galaxies within the redshift range of $0.05<z<0.07$. We exclude very low-$z$ galaxies (at $z<0.05$) because the limited size of SDSS fibre (3~arcsec) covers only a fraction of their total lights. 
\citet{2005PASP..117..227K} suggest that at least $>20\%$ of galaxy light should be covered by the fibre to reduce systematic errors from aperture effects. According to \citet{2005PASP..117..227K}, $z>0.04$ is required in the case of SDSS to ensure a covering fraction of $>20\%$ for a typical galaxy.
We note that we also apply a relatively narrow redshift range so that we can capture the same rest-frame wavelength range when we perform FIR stacking analyses (see Section~\ref{AKARI}). 

\begin{figure}
\begin{center}
\includegraphics[width=70mm]{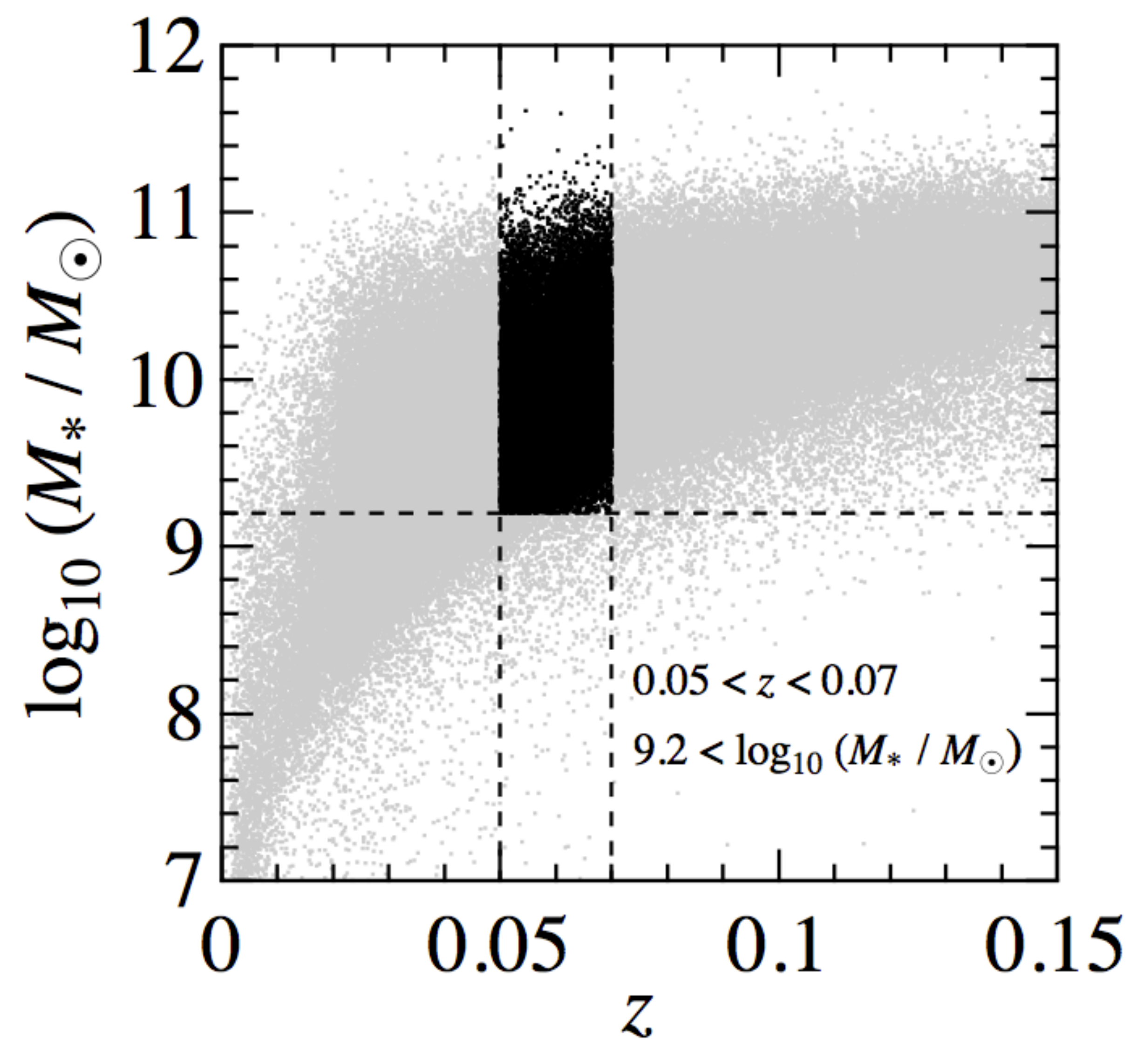}
\end{center}
\caption{Stellar mass plotted against redshift for all SDSS star-forming galaxies (grey dots). The black dots show the galaxies that we used in this paper.}
\label{MvsZ}
\end{figure}

We then apply a stellar mass cut of $9.2 < \log_{10}(M_*/M_\odot)$, considering the completeness limit of the SDSS spectroscopic survey (see Fig.~\ref{MvsZ}). Because the SDSS is magnitude limited, it is clear that the stellar mass limit depends on the redshift. 

\begin{figure}
\begin{center}
\includegraphics[width=70mm]{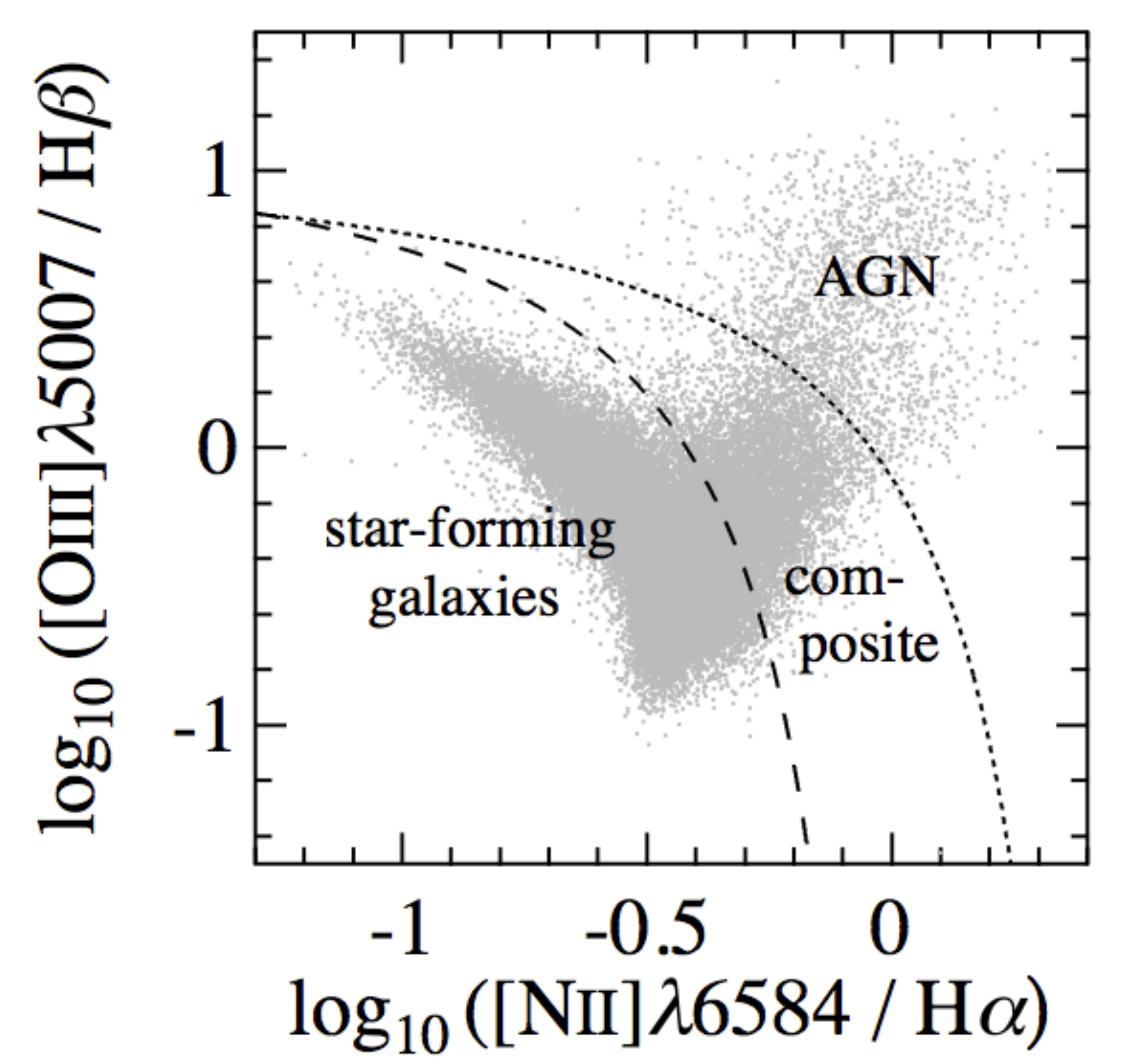}
\end{center}
\caption{The Baldwin, Phillips \& Telervich (BPT) diagram for the 42,292 SDSS galaxies with $EW_{\rm H\alpha}>$ 4 \AA\ and line flux $\rm{S/N}>3$ in $0.05<z<0.07$. The dotted and dashed lines are derived from \citet{2001ApJS..132...37K} and \citet{2003MNRAS.346.1055K}, respectively.}
\label{BPT}
\end{figure}

The main aim of this study is to investigate the environmental dependence of star-forming galaxy properties.
We select SF galaxies by applying the $EW_{\rm H\alpha}>$ 4 \AA\ and $\rm{S/N}$(H$\alpha$)$>3$ to exclude quiescent galaxies. Then, we use Baldwin, Phillips \& Telervich (BPT) diagram, which compares the $F_{\rm [OI\hspace{-.1em}I\hspace{-.1em}I]5007}/F_{\rm H\beta}$ and $F_{\rm[NI\hspace{-.1em}I]6584}/F_{\rm H\alpha}$ line flux ratio, to distinguish between star-forming galaxies and active galactic nuclei \citep{1981PASP...93....5B} (see Fig. \ref{BPT}).
For this purpose, we also require $\rm{S/N}>3$ in all the four major lines. Our final sample includes 34,249 SF galaxies to all of which local density measurements are available. 

\begin{figure*}
\begin{center}
\includegraphics[width=170mm]{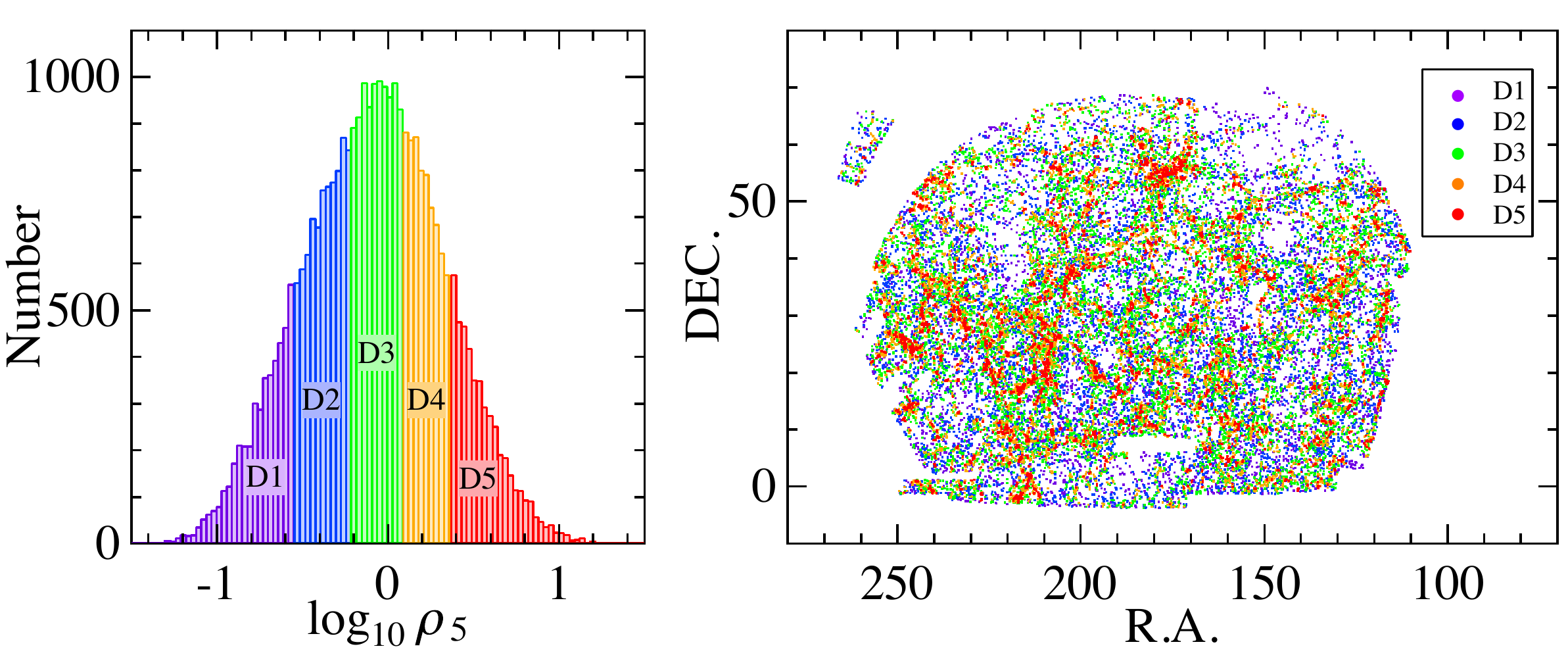}
\end{center}
\caption{(Left): The distribution of the local galaxy density ($\rho_5$) for all our SDSS star-forming galaxies. The histogram is created with the bin width of 0.05~dex. (Right): Spatial distribution of our SDSS star-forming galaxy sample on the sky. Redder symbols indicate galaxies in higher-density environments. This plot visually demonstrates the robustness of our density measurements. }
\label{RADEC}
\end{figure*}

We define the five local density bins as follows:
\begin{eqnarray}
&\rm D1&:\ \log_{10}\rho_5\leq -0.55\nonumber\\
&\rm D2&:\ -0.55<\log_{10}\rho_5\leq -0.22\nonumber\\
&\rm D3&:\ -0.22<\log_{10}\rho_5\leq 0.08\nonumber\\
&\rm D4&:\ 0.08<\log_{10}\rho_5\leq 0.36\nonumber\\
&\rm D5&:\ 0.36<\log_{10}\rho_5\nonumber
\end{eqnarray}
The size of each density bin is set to $\sim0.3$ dex for D2, D3, and D4, while D1 and D5 cover all remaining samples in lower and higher density regions, respectively (see the distribution of $\log_{10}\rho_5$ in the left panel of Fig.~\ref{RADEC}). The numbers of galaxy samples in each density bin are shown in Table~\ref{tableNsample}.

\begin{figure}
\begin{center}
\includegraphics[width=70mm]{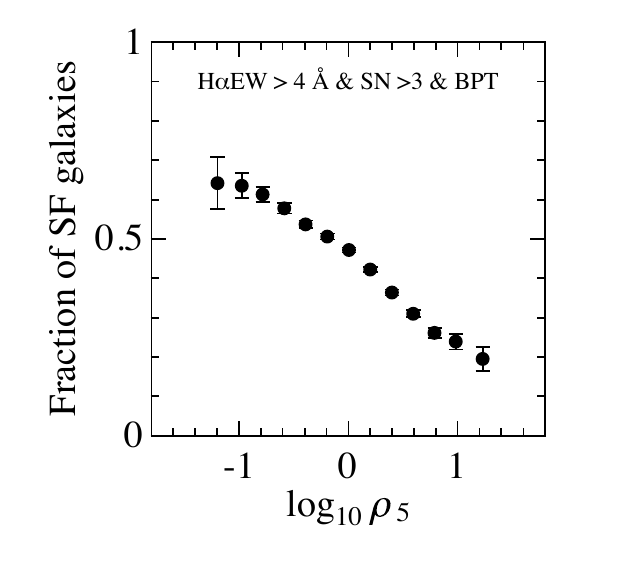}
\end{center}
\caption{Star-forming galaxy fraction as a function of the local density, showing the monotonical decrease towards high-density environment (as expected). This plot further demonstrates the validity of our density measurements.}
\label{ratioSFG}
\end{figure}

In Fig.~\ref{RADEC} (right), we show the spatial distribution of our SF galaxy samples on the sky. The redder colours indicate galaxies in higher-density environments. For example, D5 galaxies (shown with red dots) tend to be strongly clustered and align filamentary structures, while the D1 galaxies (purple dots) tend to be more widely scattered around. This result further supports the robustness of our local density measurements. We also show in Fig.~\ref{ratioSFG} the fraction of SF galaxies as a function of $\rho_5$. This plot shows a monotonical decrease of SF galaxy fraction towards higher $\rho_5$, further supporting the validity of our density measurements. 

\begin{table}
\begin{center}
\caption{The numbers of SDSS star-forming galaxy samples at $0.05 < z < 0.07$ in each environmental bin.\vspace{5mm}}
\begin{tabular}{cc} \hline
Local Density Bin & Number of Samples \\ \hline
D1 & 4767\\
D2 & 8007\\
D3 & 9628\\
D4 & 7078\\
D5 & 4769\\\hline
\end{tabular}
\label{tableNsample}
\end{center}
\end{table}

\subsubsection{Stellar mass and star formation rate}
We use the stellar mass ($M_*$) estimates from the MPA/JHU catalogue \citep{2003MNRAS.341...33K,2007ApJS..173..267S}. 
They are based on fits to the broad-band $ugriz$ photometry of the SDSS with Bayesian methodology 
following the philosophy of \citet{2003MNRAS.341...33K} and \citet{2007ApJS..173..267S}. 
The $ugriz$ broad-band magnitudes are corrected for emission lines by assuming that the relative 
contribution of emission lines to the broad-band magnitudes is the same for inside and outside the fibre,
although the effect of emission lines on broad-band magnitudes is reported to be negligibly small  
(typically $<0.1$ magnitude\footnote{http://wwwmpa.mpa-garching.mpg.de/SDSS/DR7/mass\_comp.html}).

We also use H$\alpha$-based $SFR$s calculated for the SDSS DR7 data by the MPA/JHU group (hereafter $SFR_{\rm SDSS}$). 
The calculation is based on the technique discussed in \citet{2004MNRAS.351.1151B}, 
by fitting nebular emission-line fluxes. The fits to star forming galaxies 
were performed using the \citet{2001MNRAS.323..887C} model. 
To compute the line and continuum emission from galaxies consistently, the model combines population synthesis 
and photoionization codes. The model assumes that no ionizing radiation escapes the galaxies.
The aperture correction was performed following the philosophy of \citet{2007ApJS..173..267S}.
The dust extinction correction was also done by using the $F_{\rm H\alpha}/F_{\rm H\beta}$ line flux ratio.

\subsection{AKARI}\label{AKARI}
\subsubsection{AKARI all-sky survey map}
We here describe the summary of AKARI all-sky survey data.
To investigate the environmental dependence of $SFR_{\rm IR}$ and dust 
temperature ($T_{\rm dust}$), we use the newly-released AKARI FIR 
all-sky survey maps by the Far-infrared Surveyor (FIS) \citep{2007PASJ...59S.389K} on the AKARI 
satellite \citep{2007PASJ...59S.369M}.
The data taken by FIS were pre-processed using the AKARI FIS pipeline 
tool \citep{2009ASPC..418....3Y}. The calibrations include corrections 
for non-linearity and sensitivity drifts of detectors, rejection of anomalous 
data due to high-energy particles (glitches), signal saturation, and other 
instrumental effects as well as dark-current subtraction \citep{2015PASJ...67...50D}. 
After these calibrations, the final FIS image has a pixel scale of $15''$ and 
units of surface brightness in MJy\,sr$^{-1}$. The image data are disclosed 
in FITS format files, and distributed as a number of $6.0\times6.0$ deg$^2$ 
images with ecliptic coordinate\footnote{http://www.ir.isas.jaxa.jp/AKARI/Archive/Images/FISMAP/}. 

All-sky survey was performed by AKARI in six bands centred at 9, 18, 65, 
90, 140 and 160 $\mu$m. FIS observed the four longer wavelength bands: 
i.e.\ N60, WIDE-S, WIDE-L, and N160, centred at 65, 90, 140, and 160 $\mu$m, respectively.
The point source flux detection limits (for S/N$>$5) for one scan are 2.4, 0.55, 1.4 and 6.3 Jy 
for N60, WIDE-S, WIDE-L and N160, respectively \citep{2007PASJ...59S.389K}. 
The FIS all-sky map covers $\sim$98\% of the whole sky, including the 
SDSS DR7 survey region.

Because of the limited depths of the all-sky survey, we cannot detect the 
infrared emission from individual galaxies except for very bright sources. 
We therefore decided to perform stacking analysis to study average 
properties of SF galaxies (see below).

\subsubsection{Stacking analysis}\label{stacking}
In this section, we describe the procedure of our stacking analysis.
We first split the samples into some groups according to their local density 
($\rho_5$), stellar mass, $SFR$, and $SSFR$. For each galaxy, we retrieve a 
$6.0\times6.0$ deg$^2$ map whose central position is nearest to the 
target galaxy, and create $20'\times20'$ cut out image centred at the 
position of each target galaxy. We note that we fix the y-axis directions 
of each cut-out image to the FIS scan direction.

Before stacking, we apply the following criteria to discard images 
which are not suited for the stacking analysis: 
 
\begin{enumerate}
\item Discard the images if the $20'\times20'$ cutout images protrude 
from the original 6.0 $\times$ 6.0 deg$^2$ map; i.e. sources located very close to the 
edge of $6.0\times6.0$ deg$^2$ map. 
\item Discard the images having areas whose $N_{\rm scan}$ value (the number of visit) 
is 0 or 1 (times) because the pixel values of these regions are very noisy. 
\item Discard the images that include more than 10 pixels whose values are 
smaller than $-8$ MJy\,sr$^{-1}$. This problem seems to happen when high-energy 
particles hit the detector. 
\item Discard the images having a 3 $\times$ 3 region whose total value is 
unrealistically high (with $>$300 MJy\,sr$^{-1}$ for N160 or 200 MJy\,sr$^{-1}$ 
for the others bands) at $>$30$''$ away from the image centre. 
This is most likely due to contamination of nearby bright sources. 
\end{enumerate}

\begin{figure}
\begin{center}
\includegraphics[width=80mm]{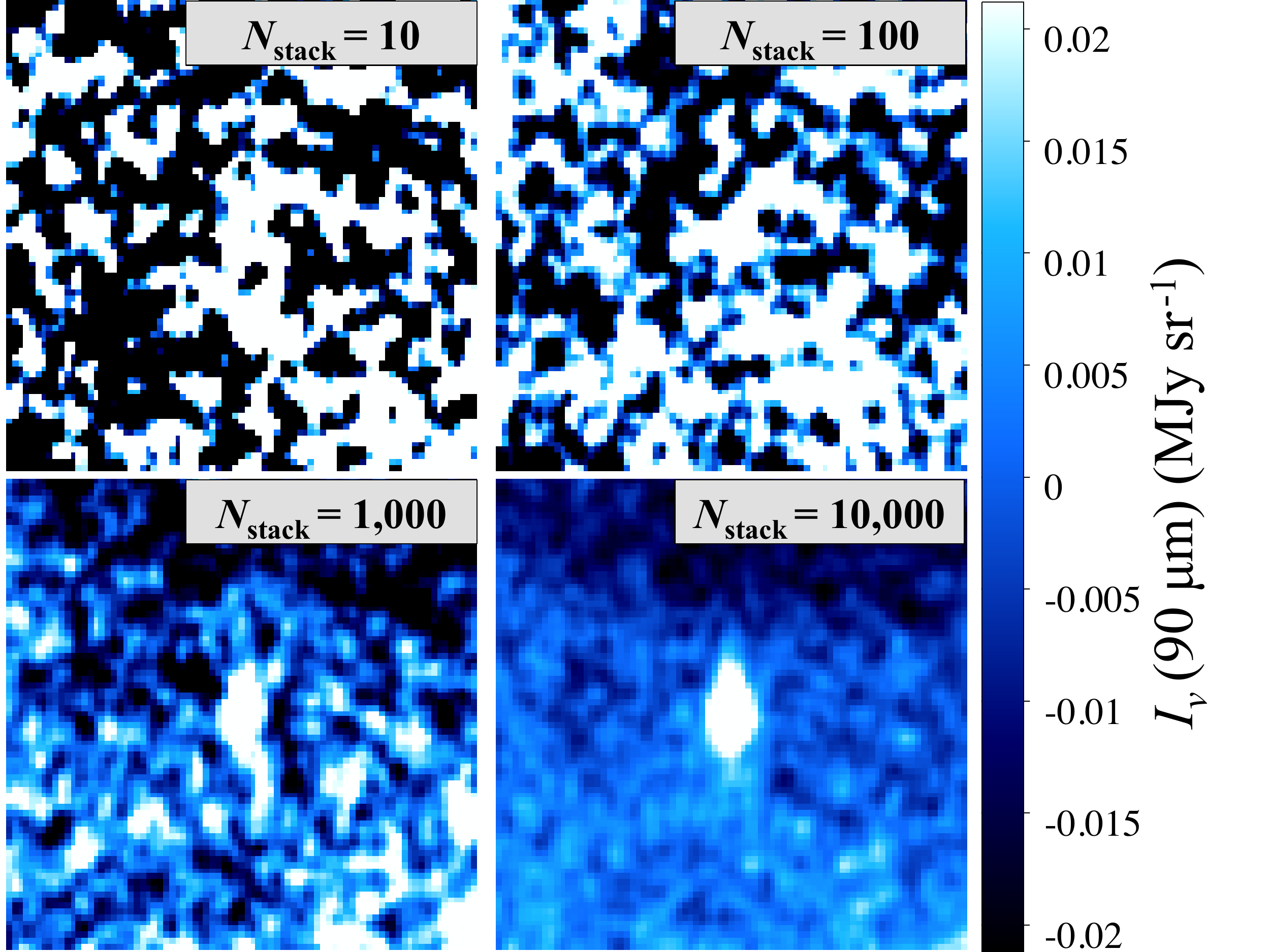}
\end{center}
\caption{Examples of the WIDE-S ``stacked'' images for galaxies with $9.2<\log_{10}M_*<9.8$, 
demonstrating the increase of signal-to-noise ratio by coadding a number of frames. 
The size of each image is $20'\times20'$. 
The numbers of coadded frames ($N_{\rm stack}$) are indicated in each panel.
The colour bar shows the scale of the intensities ($I_\nu$) at 90~$\mu$m.}
\label{ds9}
\end{figure}

\begin{figure}
\begin{center}
\includegraphics[width=80mm]{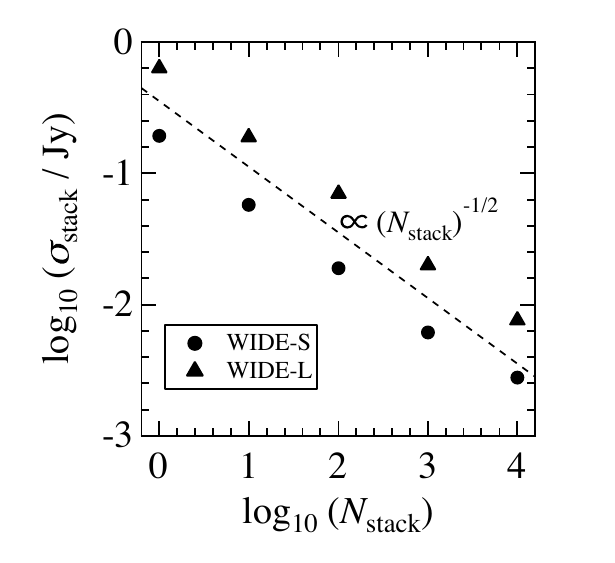}
\end{center}
\caption{The random noise levels for the stacked images for WIDE-S and WIDE-L bands 
as a function of the number of coadded frames ($N_{\rm stack}$). This plot demonstrates 
that the noise levels can be reduced with $\propto N_{\rm stack}^{-1/2}$ (see texts for details 
how we determine the noise levels of the stacked images). }
\label{sigmaStack}
\end{figure}

After applying these criteria, 
the numbers of frames are decreased by $\lesssim$ 10\%, 
and all the remaining sample are used for stacking. 
In order to derive their average flux densities of SDSS SF galaxies 
split according to stellar mass, $(S)SFR$, and environment, we perform 
average stacking for each subsample. 
We show in Fig.~\ref{ds9} how the noise levels can be reduced by 
stacking 1, 10, 100, 1,000, and 10,000 frames at the positions of 
galaxies with $9.2<\log_{10}M_*<9.8$ (for WIDE-S band data). 
This demonstrates that we can measure the average FIR fluxes of 
faint low-mass galaxies by adopting stacking analysis. 
Fig.~\ref{sigmaStack} shows this result more quantitatively; 
the noise level is reduced following $\propto N_{\rm stack}^{-1/2}$ (as expected) 
for both WIDE-S and WIDE-L data.  

We perform aperture photometry on the final stacked images.
It is suggested that there is no systematic correlation between the 
source flux densities and the PSFs \citep{2014PASJ...66...47A}, 
and so we assume that all sources have the same PSF at each band.
The full width at half maximum (FWHM) of PSFs are $63.4\pm0.2$, $77.8\pm0.2''$ and 
$88.3\pm0.9''$ at N60, WIDE-S and WIDE-L bands, respectively, hence
we can assume that our target galaxies at $0.05<z<0.07$ are point sources.
Note that the PSF at the N160 band is not determined because of the limited number of bright standard stars for this band \citep{2015PASJ...67...51T}.
The shape of the PSF at each band is anisotropic, especially for 
WIDE-S band, but we adopt simple circular aperture photometry with 
a reasonably large aperture size. We set an aperture radius to 
$90''$ and a sky background annulus area to 120--$300''$ radius,
to be consistent with the procedure shown by \citet{2015PASJ...67...51T}.

We follow the flux calibration for each band as shown by 
\citet{2015PASJ...67...51T}. The mean observed-to-expected flux 
density ratios in the case of the above aperture photometry 
parameters are $0.627\pm0.029$, $0.696\pm0.008$ 
and $0.381\pm0.043$ for the N60, WIDE-S, and WIDE-L bands, respectively. 
The calibration for N160 (160 $\mu$m) has not yet been done because 
of the limited number of bright standard stars detected at N160. 
For this reason, we do not use the 160 $\mu$m band when we derive 
$SFR_{\rm IR}$ and $T_{\rm dust}$. In addition, it is suggested that 
the stochastic heating of very small grains affects the 65 $\mu$m 
fluxes of galaxy SEDs \citep{2003ARA&A..41..241D, 2010ApJ...724L..44C}.
Therefore, we decide to compute $SFR_{\rm IR}$ and $T_{\rm dust}$ analytically 
by using the 90~$\mu$m and 140~$\mu$m fluxes.

Assuming that the flux uncertainties are dominated by random background 
noise (as demonstrated by our analyses in Fig.~\ref{sigmaStack}), 
we determine the flux errors with the following procedure. 
We first select $N_{\rm stack}$ random positions on the FIS map, and 
stack them. We then perform aperture photometry on this stacked 
image in the same way as above. We repeat this process 200 times, 
and we take the standard deviation of 200 flux densities as our 
1$\sigma$ flux uncertainties. We note that the uncertainties for 
$SFR_{\rm IR}$ and $T_{\rm dust}$ presented in this work simply 
reflect the photometric errors estimated here. 

\section{SPECIFIC STAR FORMATION RATE VERSUS ENVIRONMENT}\label{MS}
\subsection{Deriving total infrared luminosities ($L_{\rm IR}$)}
Interstellar dust absorbs and scatters the UV light from O/B-type stars and re-emit it in far infrared (FIR). 
Therefore, the total infrared luminosities ($L_{\rm IR}$) are known to be a good tracer for dust-enshrouded 
star formation activities \citep{1998ARA&A..36..189K,2001ApJ...556..562C}.

In this section, we summarize the method for deriving $SFR_{\rm IR}$.
Firstly, we derive $L_{\rm IR}$ by using the AKARI FIR photometry 
at WIDE-S and WIDE-L bands. Following \citet{2010A&A...514A...4T}, 
we here define $L_{\rm AKARI}^{2{\rm bands}}$ as the following:
\begin{eqnarray}
L_{\rm AKARI}^{2{\rm bands}}=\Delta\nu_{90\mu{\rm m}}L_{\nu}(90\mu{\rm m})+\Delta\nu_{140\mu{\rm m}}L_{\nu}(140\mu{\rm m}),
\end{eqnarray}
where $L_{\nu}(90\mu{\rm m})$ and $L_{\nu}(140\mu{\rm m})$ are the luminosity densities at WIDE-S and WIDE-L respectively, and
\begin{eqnarray}
\Delta\nu_{90\mu{\rm m}}=1.47\times10^{12} (\rm Hz)\\
\Delta\nu_{140\mu{\rm m}}=0.831\times10^{12} (\rm Hz)
\end{eqnarray}
are the band widths for WIDE-S and WIDE-L, respectively \citep{2008PASJ...60S.477H}.
We derive $L_{\nu}(90\mu{\rm m})$ and $L_{\nu}(140\mu{\rm m})$ as follows:
\begin{eqnarray}
L_{\nu}(90\mu{\rm m})=4\pi d_{\rm L}^2\frac{F_{\nu}(90\mu{\rm m})}{1+z},\\
L_{\nu}(140\mu{\rm m})=4\pi d_{\rm L}^2\frac{F_{\nu}(140\mu{\rm m})}{1+z},
\end{eqnarray}
where  $F_{\nu}(90\mu{\rm m})$ and $F_{\nu}(140\mu{\rm m})$ are the observed flux densities 
at WIDE-S and WIDE-L bands, respectively, and $d_{\rm L}$ is the luminosity distance to 
each galaxy. Then, $L_{\rm IR}$ is derived from $L_{\rm AKARI}^{2{\rm bands}}$ with the following 
equation shown by \citet{2010A&A...514A...4T}:
\begin{eqnarray}
\log_{10}(L_{\rm IR}/L_\odot)=0.964\log_{10}L_{\rm AKARI}^{2{\rm bands}}+0.814.
\end{eqnarray} 
To keep consistency with measurements for SDSS H$\alpha$-based $SFR$s ($SFR_{\rm SDSS}$), 
we derive IR-based $SFR$ with the relation from \citet{1998ARA&A..36..189K} assuming 
Kroupa IMF:
\begin{eqnarray}
SFR_{\rm IR}\ (M_\odot\,{\rm yr}^{-1})=3.1\times10^{-44}L_{\rm IR}\ (\rm erg\,s^{-1}).
\end{eqnarray} 

\begin{figure}
\begin{center}
\includegraphics[width=80mm]{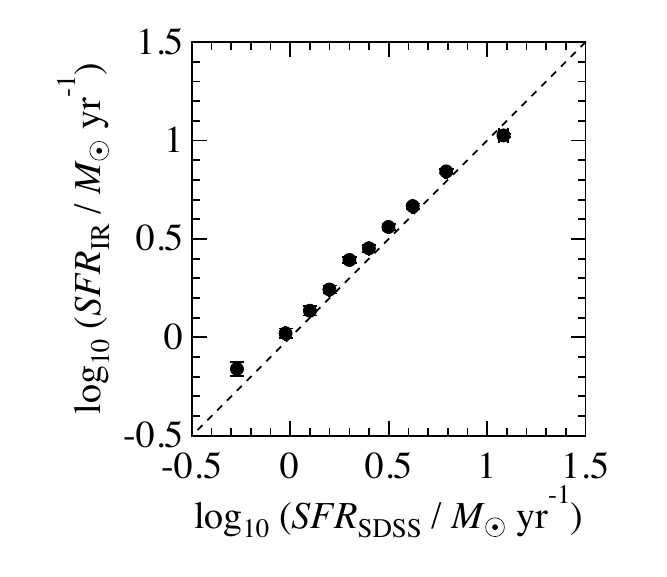}
\end{center}
\caption{Comparison between the IR-based $SFR$s ($SFR_{\rm IR}$) and the H$\alpha$-based $SFR$s ($SFR_{\rm SDSS}$) for SDSS star-forming galaxy samples divided into 10 $SFR_{\rm SDSS}$ bins. The dashed line corresponds to: $SFR_{\rm IR}$=$SFR_{\rm SDSS}$. This plot demonstrates that the two independent measurements agree with each other very well.}
\label{SFR_IRvsSDSS}
\end{figure}

Here we do not take into account the effect of $k$-correction because 
the WIDE-S and WIDE-L bands are wide enough, and the effects can be negligible 
when measuring $L_{\rm IR}$ in the redshift range we are considering.
Indeed, it is demonstrated that $SFR_{\rm IR}$s derived with this method 
show good agreement with $SFR$s derived from dust-corrected H$\alpha$ luminosities 
for $z< 0.1$ galaxies \citep{2015MNRAS.453..879K}. 
As a further check, we also plot in Fig.~\ref{SFR_IRvsSDSS} our data points 
(from stacking) on the $SFR_{\rm IR}$--$SFR_{\rm SDSS}$ plane. 
We here further divide our SF galaxy sample into 10 $SFR_{\rm SDSS}$ bins, 
and perform the stacking analyses for each subsample to derive their $SFR_{\rm IR}$. 
It is found that $SFR_{\rm IR}$ are in good agreement with $SFR_{\rm SDSS}$
over wide luminosity range. This result supports the validity of our procedure 
for deriving $L_{\rm IR}$.

\subsection{Environmental dependence of $SSFR$ for SF galaxies}\label{SSFR_environment}

Our primary goal is to investigate the environmental dependence of $T_{\rm dust}$.
As shown by \citet{2014A&A...561A..86M}, there exists a strong correlation between $T_{\rm dust}$ and $SSFR$.
Therefore, we must distinguish the effects of environment and $SSFR$. 
For this purpose, we examine the environmental dependence of IR-based $SSFR$ ($SSFR_{\rm IR}$).

The unit of $SSFR$ is inverse of time, and so $SSFR$ indicates the time-scale of star formation.
We calculate both IR- and H$\alpha$-based $SSFR$ (hereafter $SSFR_{\rm IR}$ and $SSFR_{\rm SDSS}$).
In order to perform fair comparison between $SSFR_{\rm IR}$ (from stacking) and $SSFR_{\rm SDSS}$, 
we derive the mean $SSFR_{\rm SDSS}$ 
by dividing the mean $SFR$ by the mean $M_*$ rather than simply averaging the $SFR$/$M_*$ of 
individual galaxies: i.e.
\begin{eqnarray}
\langle SSFR\rangle=\frac{\langle SFR\rangle}{\langle M_*\rangle}\label{SSFR}\\
\nonumber\neq\left\langle\frac{SFR}{M_*}\right\rangle
\end{eqnarray}

\begin{figure}
\begin{center}
\includegraphics[width=85mm]{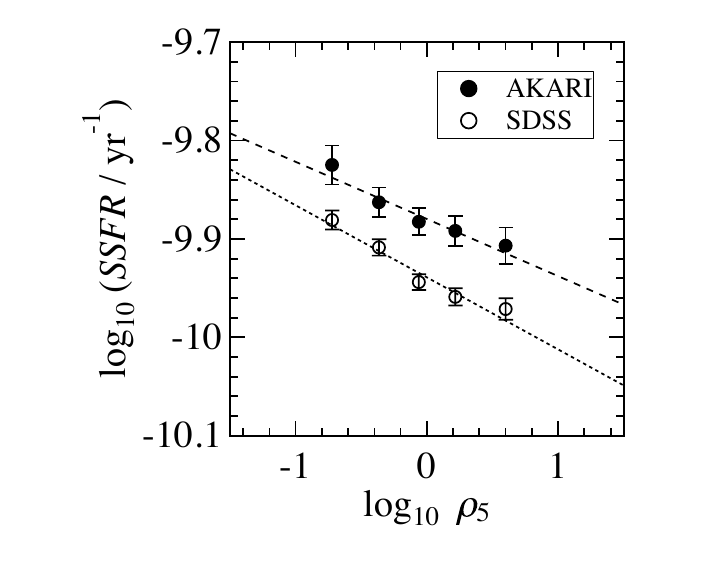}
\end{center}
\caption{Dependence of the $SSFR_{\rm IR}$ (filled circles) and $SSFR_{\rm SDSS}$ (open circles) on the local density. 
The dashed and dotted lines are the best fitted lines corresponding to: $\log_{10}SSFR_{\rm IR}=-(9.880\pm0.007)-(0.06\pm0.02)\times\log_{10}\rho_5$ and $\log_{10}SSFR_{\rm SDSS}=-(9.939\pm0.004)-(0.07\pm0.01)\times\log_{10}\rho_5$, respectively. We note that the results shown in this diagram are for star-forming galaxies only. }
\label{SSFR_IR_vs_rho5}
\end{figure}

In Fig.~\ref{SSFR_IR_vs_rho5}, we show the average $SSFR_{\rm IR}$ and $SSFR_{\rm SDSS}$ derived for each local density bin. The best fitted lines to the data points are: 
\begin{eqnarray}
&\log_{10}(SSFR_{\rm IR}/yr^{-1})=&\\
&-(9.880\pm0.007)-(0.06\pm0.02)\times\log_{10}\rho_5,&\nonumber
\end{eqnarray} 
and  
\begin{eqnarray}
&\log_{10}(SSFR_{\rm SDSS}/yr^{-1})=&\\
&-(9.939\pm0.004)-(0.07\pm0.01)\times\log_{10}\rho_5.&\nonumber
\end{eqnarray} 
The best fitted lines show negative slope, indicating that both $SSFR_{\rm IR}$ and $SSFR_{\rm SDSS}$ 
monotonically decrease with increasing $\rho_5$. We note that our sample includes 
{\it only SF galaxies}. Recent studies reported 
lack of $SFR$--density relation amongst SF galaxies \citep{2004MNRAS.348.1355B,2012MNRAS.423.3679W},
but our result sugggests that there is small, but significant environmental dependence 
of $SSFR$ for SF galaxies. 
Because the scatter of the SF main sequence is reported to be small ($\sim$0.3-dex;  
e.g.\ \citealt{2007A&A...468...33E}), the $\sim$0.1-dex environmental dependence 
of $SSFR$ would not be negligible.
We need to consider this effect when we discuss environmental dependence of $T_{\rm dust}$
(see Section~\ref{Td_environment}).

We comment that the $SSFR_{\rm IR}$ tends to be slightly higher than the $SSFR_{\rm SDSS}$ 
($\sim$0.05-dex level) as can be seen in Fig.~\ref{SSFR_IR_vs_rho5}. 
Actually, this small offset can also be seen in Fig.~\ref{SFR_IRvsSDSS}. 
The reason is unclear, but this small difference between 
$SFR_{\rm SDSS}$ and $SFR_{\rm IR}$ has no significant effect on our main conclusion at all. 

Our results are consistent with several recent studies which reported the decline 
in $SSFR$ of SF galaxies in high-density environments. For example, \citet{2013ApJ...775..126H} found that the 
$SSFR$ of SF galaxies in clusters is systematically ($\sim28$\%) lower than their 
counterparts in the field, which is qualitatively consistent with our results. 
\citet{2016MNRAS.tmp...95F} also found that $SSFR$s of late type galaxies in the Coma cluster 
decrease monotonically with increasing local galaxy number density by using the {\it Herschel} 
Astrophysical Terahertz Large Area Survey (H-ATLAS) data. 
The environmental difference in $SSFR$ reported by \citet{2016MNRAS.tmp...95F} is 
$\sim1$-dex level at maximum, which is much larger than our result. 
Although the exact reason of this difference is not clear, a potential reason 
would be the different environmental range considered in their studies and our current work. 
We also note that their late-type galaxy sample is morphologically selected, while our 
star-forming galaxy sample is selected with the emission line properties.

\section{Dust temperature versus environment}\label{Td}
\subsection{Derivation of $T_{\rm dust}$}\label{DerivationTd}
We derive $T_{\rm dust}$ using the FIR multi-band photometry obtained by our stacking analysis (see Section~\ref{stacking}).
In order to estimate the average $T_{\rm dust}$ for each subsample of SDSS SF galaxies, we here assume 
a single modified blackbody function:
\begin{eqnarray}
\nonumber F_\nu&=&C\nu^\beta B_\nu(T_{\rm dust})\label{Fnu}\\\nonumber
&=&C\frac{\nu^{3+\beta}}{\exp(h\nu/kT_{\rm dust})-1}\\
&=&C\frac{(c/\lambda)^{3+\beta}}{\exp[(hc/\lambda)/kT_{\rm dust}]-1}\label{MMB},
\end{eqnarray} 
where $F_\nu$ is the flux density, $\beta$ is the dust emissivity spectral index, 
$C$ is a free-floating normalization factor, and $B_\nu$ is the Planck blackbody 
radiation function. 
Following many other works studying dust temperature in galaxies, we fix $\beta=1.5$ in this study.

As discussed in Sec \ref{AKARI}, we calculate the dust temperature using the flux densities 
at 90 and 140 $\mu$m, which straddle the peak of FIR SEDs of SF galaxies. 
Using equation (\ref{Fnu}), the ratio of the flux densities in the two bands can be written as:
\begin{eqnarray}
\frac{F_{\nu_2}}{F_{\nu_1}}=\frac{\nu_2^{3+\beta}[\exp(h\nu_1/kT_{\rm dust})-1]}{\nu_1^{3+\beta}[\exp(h\nu_2/kT_{\rm dust})-1]},\label{nonapprox}
\end{eqnarray} 
where $F_{\nu_1}$ and $F_{\nu_2}$ are the flux densities at 90 and 140 $\mu$m, respectively. 
This equation can then be approximated as: 
\begin{eqnarray}
\ln\frac{F_{\nu_2}}{F_{\nu_1}}\approx (3+\beta)\ln\frac{\nu_2}{\nu_1}+\frac{h}{kT}(\nu_1-\nu_2).\label{approx}
\end{eqnarray} 
We checked that the difference between the $F_{\nu_2}/F_{\nu_1}$ value from the equation (\ref{nonapprox}) and (\ref{approx}) for $\beta=1.5$ and $10<T_{\rm dust}<30$ is only 2\% at maximum, which does not affect our results.
Therefore, we can analytically derive $T_{\rm dust}$ with the following equation: 

\begin{eqnarray}
\nonumber T_{\rm dust}\approx\frac{h(\nu_1-\nu_2)/k}{\ln\frac{F_{\nu_2}}{F_{\nu_1}}-(3+\beta)\ln\frac{\nu_2}{\nu_1}}\\
=\frac{\frac{hc}{k}(\frac{1}{\lambda_1}-\frac{1}{\lambda_2})}{\ln\frac{F_{\nu_2}}{F_{\nu_1}}-(3+\beta)\ln\frac{\lambda_1}{\lambda_2}}\label{eqTdust}.
\end{eqnarray}
We believe that this analytical approach is advantageous in the sense that 
it is always reproducible. One thing we should note is that the assumption 
of $\beta$ value can affect the measurement of $T_{\rm dust}$:
e.g.\ if we assume $\beta=2.0$, the resultant $T_{\rm dust}$ becomes lower (typically 
by $\sim2$ K) than those derived under the assumption of $\beta=1.5$. 
Therefore the {\it absolute} values of $T_{\rm dust}$ should be interpreted with care, 
but we stress that our {\it internal} comparisons (i.e.\ stellar mass, $SFR$, or environmental dependence 
of $T_{\rm dust}$) shown in this paper would not be affected by this uncertainty.

Following many other works studying 
dust temperature in galaxies, we assume the constant $\beta$ value when studying the environmental effects 
on $T_{\rm dust}$.

\subsection{$T_{\rm dust}$ as a function of galaxy properties}\label{Td_SSFR}
Before investigating the environmental effects on $T_{\rm dust}$, we examine 
the dependence of $T_{\rm dust}$ on various galaxy properties (e.g.\ $M_*$, $SFR$, and $SSFR$). 
We here divide the full SDSS SF galaxy sample into ten $M_*$/$SFR_{\rm SDSS}$/$SSFR_{\rm SDSS}$ bins, 
and perform FIR stacking analysis for each subsample. 

\begin{figure*}
\begin{center}
\includegraphics[width=180mm]{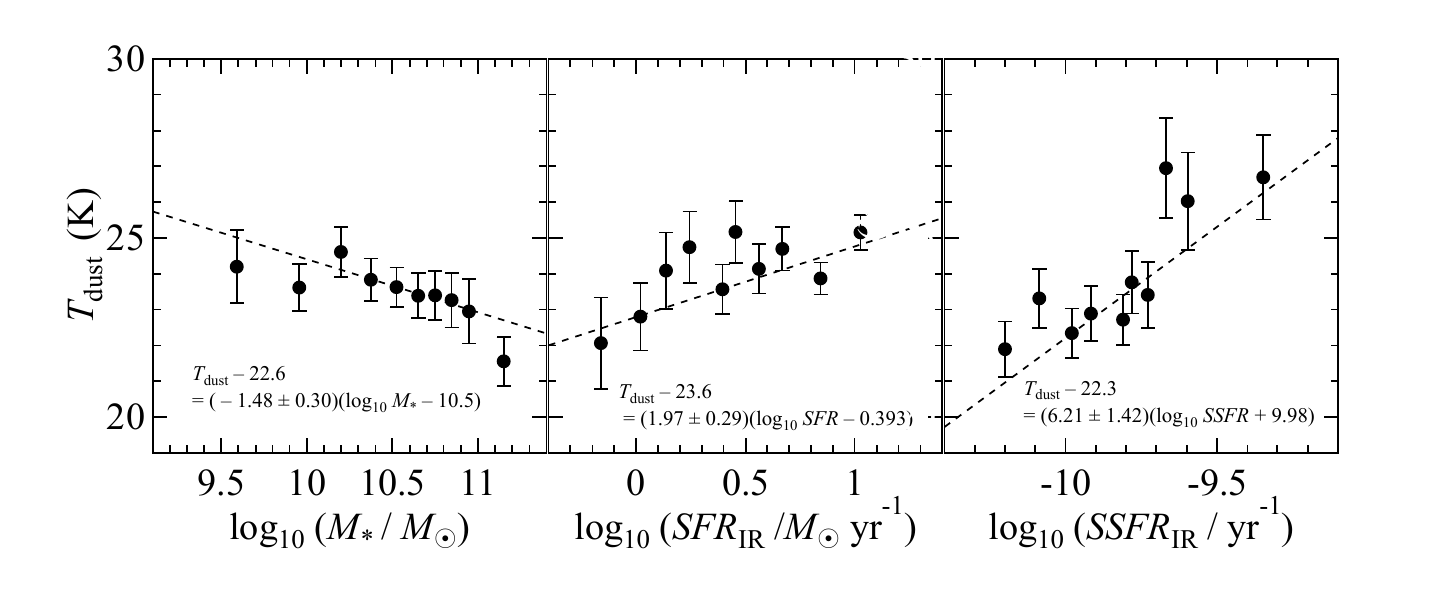}
\end{center}
\caption{Average $T_{\rm dust}$ of SF galaxies as a function of $M_*$ (left), $SFR$ (middle), and $SSFR$ (right). 
In each panel, we split the sample into ten subsamples and perform FIR stacking analysis, 
to investigate the trend of $T_{\rm dust}$ with these physical properties of galaxies. 
The dashed line indicates the line fitting result in the case we adopt a pivot point of $\log_{10}M_*=10.5$, $\log_{10} SFR =0.393$ and $\log_{10} SSFR = -9.98$.
Note that the results do not change much even if we performed the line fits without any pivot points.
}
\label{Td_vs_sSFR_SFR_M}
\end{figure*}

In the left and middle panel of Fig.~\ref{Td_vs_sSFR_SFR_M}, we plot $T_{\rm dust}$ 
as a function of $M_*$ and $SFR$, respectively. It can be seen that $T_{\rm dust}$ 
increases with $SFR$, while $T_{\rm dust}$ decreases with increasing $M_*$. 
We note that we use SDSS-based $SFR$/$SSFR$ to split the sample, but the plotted
date points show IR-based measurements from stacking analyses. 
We note that the $T_{\rm dust}$--$SFR$ correlation reported here is equivalent to the $T_{\rm dust}$--$L_{\rm IR}$ correlation shown by many recent studies (e.g. \citealt{2002ApJ...570..470T,2003MNRAS.338..733B,2010MNRAS.409...75H}).

In the right panel of Fig.~\ref{Td_vs_sSFR_SFR_M}, we plot $T_{\rm dust}$ 
against $SSFR_{\rm IR}$; we here divide the sample into ten $SSFR_{\rm SDSS}$ 
bins and perform stacking analysis in the same way as above. 
The dashed line shows the result from best line fit.
The increasing trend of $T_{\rm dust}$ with $SSFR$ is consistent with the results shown by \citet{2014A&A...561A..86M}, 
although our derived $T_{\rm dust}$ values tend to be slightly lower than those of \citet{2014A&A...561A..86M} by $\sim$1--2~K.
We here note again that most of our discussions presented in this paper are based on the {\it relative} comparison between 
our subsamples, and are not affected by the absolute value of $T_{\rm dust}$.

It is clear that there is a strong positive correlation between dust temperature and $SSFR_{\rm IR}$. 
Because $SSFR$ reflects the strength of UV radiation fields due to young massive stars in galaxies and the UV radiation heats up the dust content,
it is expected that galaxies with higher $SSFR$ tend to have warmer dust temperatures, whereas galaxies with lower $SSFR$ should 
have colder dust temperatures \citep{2014A&A...561A..86M}.

Fig.~\ref{Td_vs_sSFR_SFR_M} suggests that $T_{\rm dust}$ is most strongly correlated with $SSFR$,
compared with $M_*$ or $SFR$. This is not surprising because the $T_{\rm dust}$--$SSFR$ relation reflects 
the negative correlation between $T_{\rm dust}$ and $M_*$, as well as the positive correlation between 
$T_{\rm dust}$ and $SFR$. 
We need to consider this strong dependence of $T_{\rm dust}$ on $SSFR$ when we discuss 
environmental dependence of $T_{\rm dust}$.

\subsection{Environmental dependence of $T_{\rm dust}$}\label{Td_environment}
This section presents the main results of this paper: 
i.e.\ the environmental dependence of $T_{\rm dust}$.
We note that the combination of the wide-field coverage of 
SDSS and its entire coverage by the AKARI all-sky survey map 
allows us to perform the first systematic study of dust temperature 
of SF galaxies as a function of environment. 

\begin{figure*}
\begin{center}
\includegraphics[width=140mm]{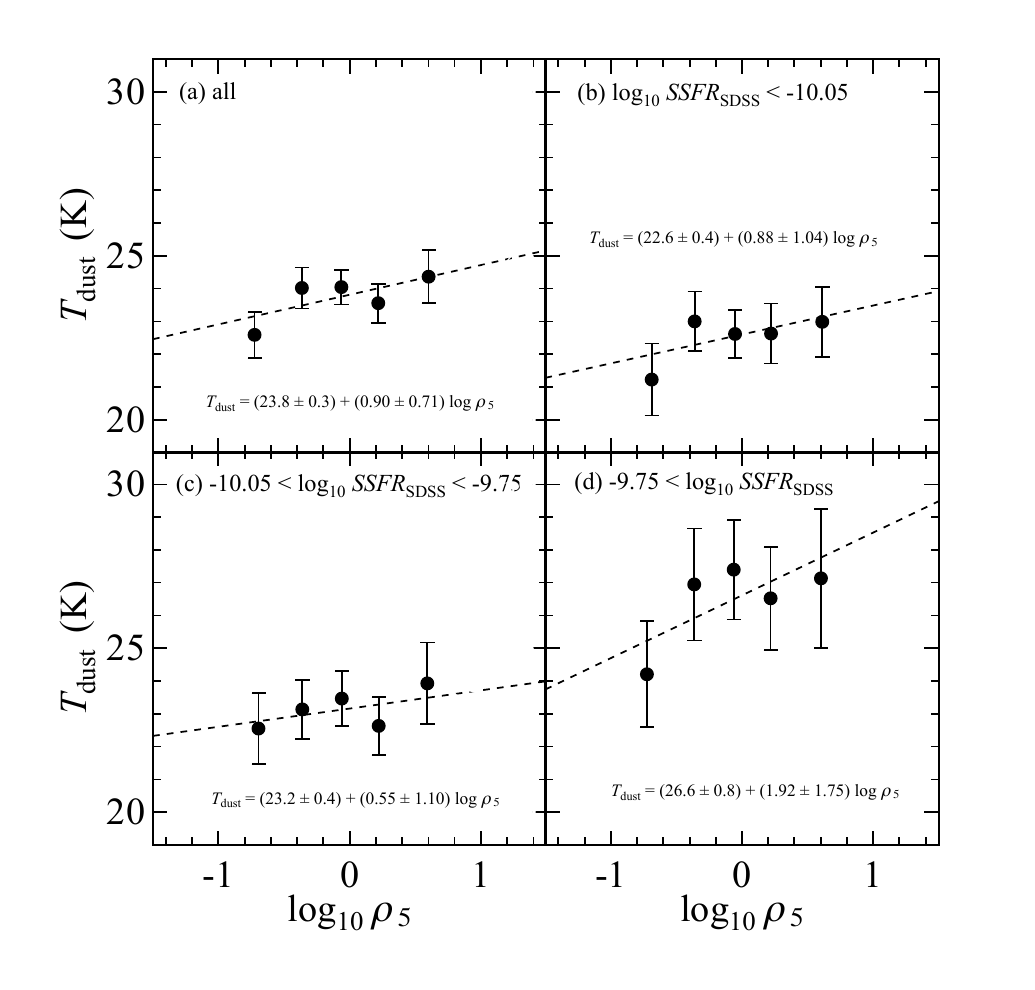}
\end{center}
\caption{(a): Dependence of $T_{\rm dust}$ on the local galaxy density ($\rho_5$) for all the sample. (b)--(d): The same plot for three $SSFR_{\rm SDSS}$ bins. In all these panels, $T_{\rm dust}$ are derived from the 90 and 140 $\mu$m data (with equation (\ref{eqTdust})). 
The definition of the $SSFR_{\rm SDSS}$ range is indicated in each panel. The dashed lines correspond to the best-fit line.}
\label{Td_vs_rho5_all_sSFR}
\end{figure*}

We recall that we defined the five environment bins (D1--D5) based on the local density of galaxies (see Section~\ref{def_environment}).
As a first step, we show in Fig.~\ref{Td_vs_rho5_all_sSFR}(a) the $T_{\rm dust}$ derived for the D1--D5 samples. 
It would be interesting to note that $T_{\rm dust}$ mildly {\it increases} with increasing environmental density. 
The best fit line in the panel (a) derived with all the star-forming galaxies corresponds to:
\begin{eqnarray}
T_{\rm dust}({\rm K}) = (23.8\pm0.3)+(0.90\pm0.71)\times\log_{10}\rho_5.
\end{eqnarray}

It turns out that the slope of this line is significantly positive (at $1.5\sigma$ level) with the probability 
of $\sim$80\%. In Section~\ref{SSFR_environment}, we reported that $SSFR$ of SF galaxies {\it decrease} with 
increasing local density. On the other hand, we also found that there is an {\it increasing} 
trend of $T_{\rm dust}$ with $SSFR$ (see Section~\ref{Td_SSFR}). By combining these two results, it is expected 
that $T_{\rm dust}$ should {\it decreases} with density---but the data suggests an opposite trend. 
We admit that the increasing trend is mostly driven by the data points of the lowest-density bin (D1),
but we verify that our results are unchanged even when we split the sample into 10 environmental bins
(although the error-bars of individual data points become larger in this case).

To investigate this trend more in detail, 
we further divide the D1--D5 sample into three $SSFR_{\rm SDSS}$ bins and 
perform the same stacking analyses to derive $T_{\rm dust}$. 
The results are shown in the panels (b)--(d) of Fig.~\ref{Td_vs_rho5_all_sSFR}.
The error bars are clearly larger because of the smaller sample size, in particular 
for the panel-(d) because galaxies with higher $SSFR_{\rm SDSS}$ tend to have lower 
stellar mass or lower luminosity. Although the statistics is poor, 
it is notable that the same increasing trend of $T_{\rm dust}$--$\rho_5$ relation 
can be seen in all the cases. 

\begin{figure*}
\begin{center}
\includegraphics[width=150mm]{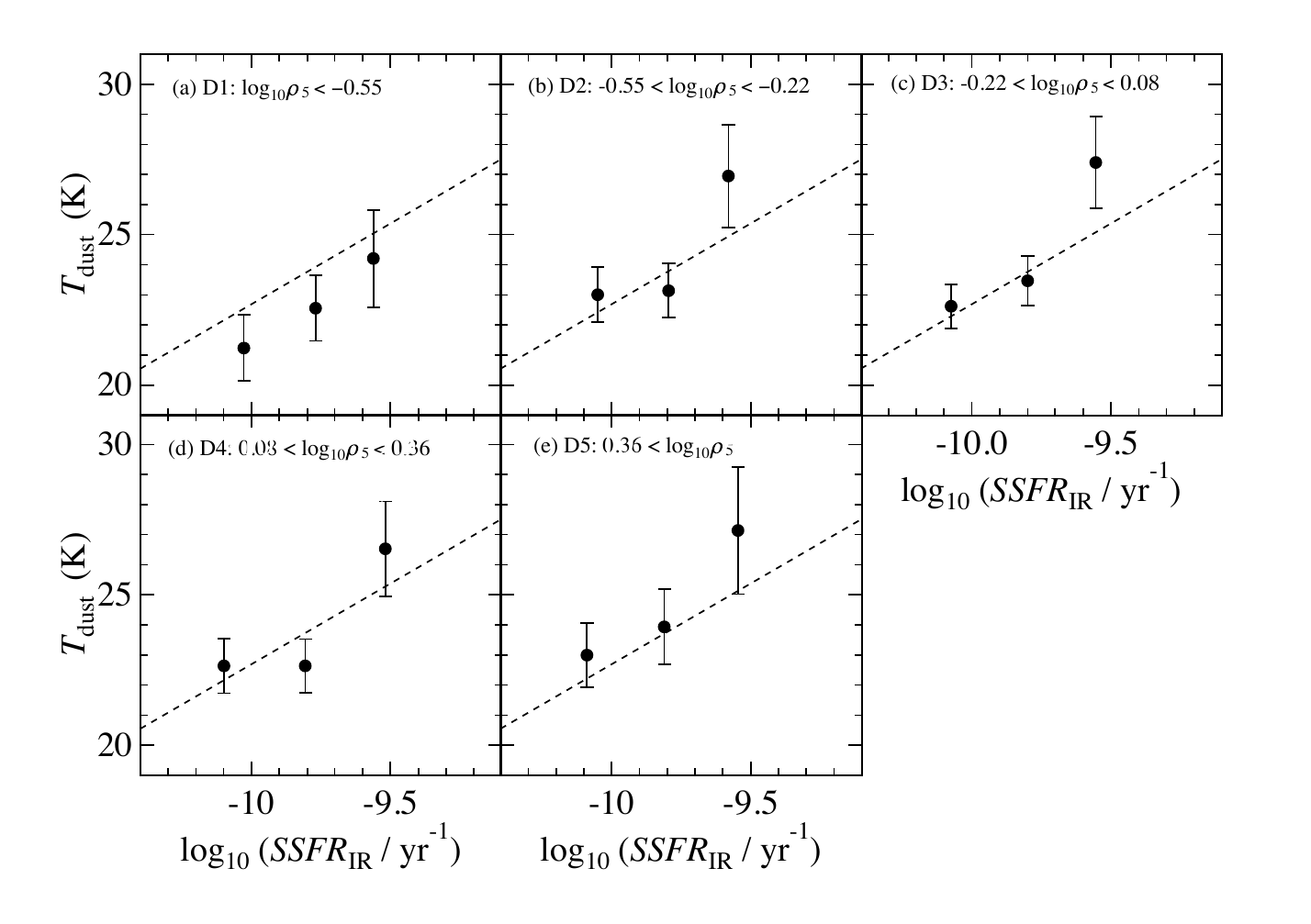}
\end{center}
\caption{Dependence of $T_{\rm dust}$ on $SSFR_{\rm IR}$ at fixed local galaxy number density (D1--D5 sample). 
The dashed line in each panel corresponds to the best-fit line in the right panel of Fig.~\ref{Td_vs_sSFR_SFR_M}.
The clear $T_{\rm dust}$--$SSFR$ correlation can be seen in all environments. 
}
\label{Td_vs_sSFRrho5}
\end{figure*}

\begin{figure*}
\begin{center}
\includegraphics[width=140mm]{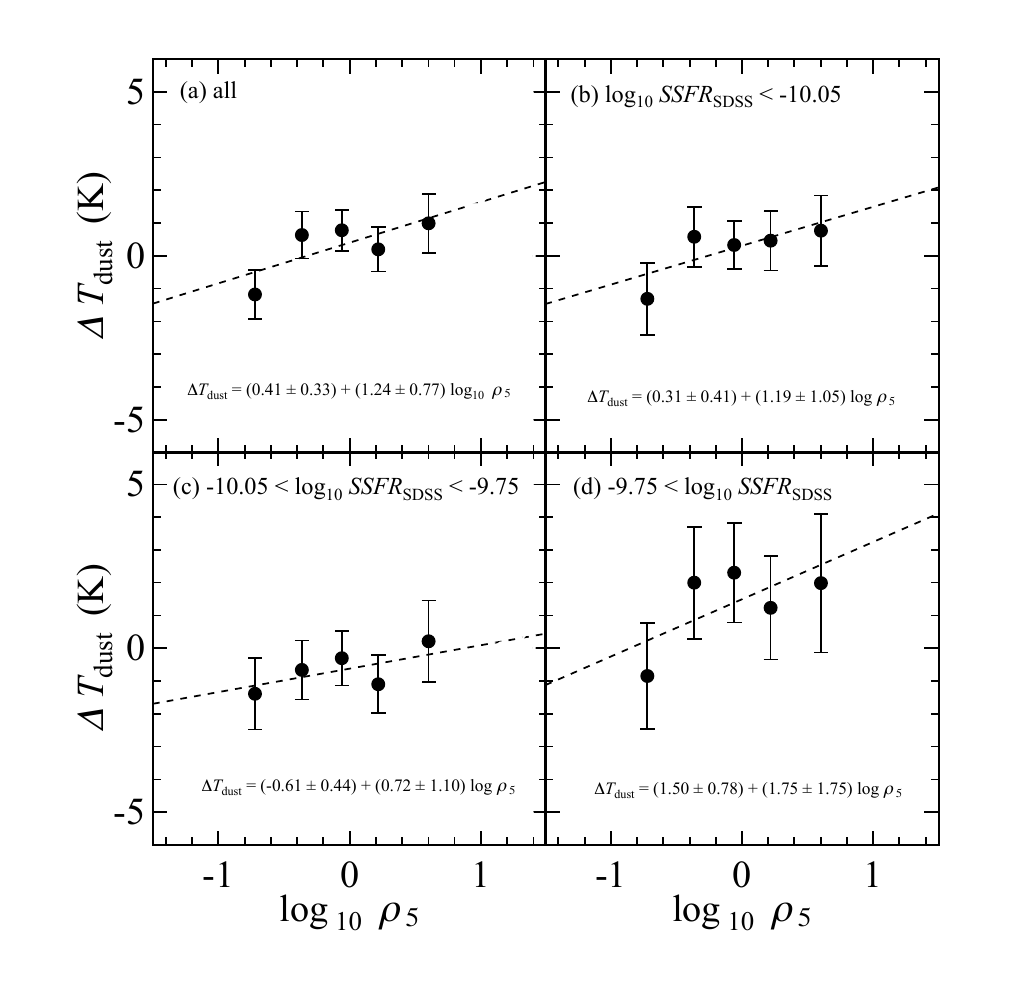}
\end{center}
\caption{(a): Environmental dependence of the average offset values ($\Delta T_{\rm dust}$) from 
the best fit $T_{\rm dust}$--$SSFR_{\rm IR}$ derived with all star-forming galaxies (Fig.~\ref{Td_vs_sSFR_SFR_M}).
(b)--(d): the same plot in each $SSFR_{\rm SDSS}$ bin. The dashed lines correspond to the best line fits.}
\label{Td_vs_sSFR_rho5_offset}
\end{figure*}

We also examine the dependence of the $T_{\rm dust}$--$SSFR_{\rm IR}$ relation on $\rho_5$. 
In each panel of Fig.~\ref{Td_vs_sSFRrho5}, we plot $T_{\rm dust}$ against $SSFR_{\rm IR}$ 
at fixed environment and compare them with the best-fit $T_{\rm dust}$--$SSFR_{\rm IR}$ relation
derived for all SF galaxies (i.e.\ dashed line in the right panel of Fig.~\ref{Td_vs_sSFR_SFR_M}).
Fig.~\ref{Td_vs_sSFRrho5} is equivalent to the panel (b)--(d) of Fig.~\ref{Td_vs_rho5_all_sSFR},
but we can now confirm that the $T_{\rm dust}$--$SSFR$ correlation is in place in all the environments. 
In other words, we need to remove the influence of $SSFR$ to 
discuss the environmental dependence of $T_{\rm dust}$ more robustly. 
For this purpose, we here define the offset value ($\Delta T_{\rm dust}$) of the data points from the best-fitted $T_{\rm dust}$--$SSFR_{\rm IR}$ line (see Fig.~\ref{Td_vs_sSFRrho5}) as follows:
\begin{eqnarray}
\Delta T_{\rm dust}=T_{\rm dust}-T_{\rm dust,exp} ,
\end{eqnarray}
where $T_{\rm dust,exp}$ is the expected $T_{\rm dust}$ estimated from the best fitted $T_{\rm dust}$--$SSFR$ relation. 
In Fig.~\ref{Td_vs_sSFR_rho5_offset}, we show $\Delta T_{\rm dust}$ for each environment.
The dashed lines in this plot show the best-fit lines.
The best fit line in the panel (a) derived with all the galaxies corresponds to:
 \begin{eqnarray}
\Delta T_{\rm dust}({\rm K}) = (0.41\pm0.33)+(1.24\pm0.77)\times\log_{10}\rho_5.
 \end{eqnarray}
It should be noted that there still remains a positive correlation between 
$\Delta T_{\rm dust}$ and $\rho_5$. 
Interestingly, we find similar trends even when we further divide D1--D5 sample into 
three $SSFR_{\rm SDSS}$ bins (see Fig.~\ref{Td_vs_sSFR_rho5_offset} (b)--(d)).
Our results suggest that $T_{\rm dust}$ increases with $\rho_5$ from low to high-density 
environment at fixed $SSFR$, and therefore 
the environmental effects on $T_{\rm dust}$ of SF galaxies should work 
over wide environmental range, and are not necessarily cluster-specific mechanisms. 
We note, however, that the average environmental difference reported here is only 
$\sim$3 K at maximum over the $\sim$2-dex local density range (D1--D5). 
Therefore, our conclusion is that the environmental impacts on dust temperature in SF galaxies 
are much milder than that of $SSFR$, but we suggest that the environment should have some impacts 
on dust temperature in SF galaxies.

\subsection{Discussion}

Dust temperature in galaxies is expected to provide us with an important information 
on the geometry of star formation taking place inside the galaxies (e.g. \citealt{2011A&A...533A.119E}).
In this paper, we reported a marginal trend that $T_{\rm dust}$ (as well as $\Delta T_{\rm dust}$) 
increases with environmental density ($\rho_5$).

The environmental impacts on dust properties of galaxies have not yet been studied well, 
but a few recent studies argue possible relationship between dust temperature and environment. 
\citet{2012ApJ...756..106R} found that some galaxies (with $\log_{10}(L_{\rm IR}/L_\odot)<11$)   
in the Bullet Cluster at $z\sim0.3$ have {\it warmer} dust temperature (by $\sim$7~K) 
than field galaxies with same luminosity using {\it Herschel} data.
They suggested that dust stripping is the responsible mechanism for the unusually warm 
dust temperature, as the stripping effect can more easily strip cold dust from 
outskirts of galaxies. However, it is also true that $\sim90$\% of their star-forming cluster 
galaxies have dust temperatures similar to field galaxies with comparable $L_{\rm IR}$.
We expect that the moderate rise of $T_{\rm dust}$ in higher-density environments reported by our analyses 
is qualitatively consistent with the results shown by \citet{2012ApJ...756..106R}.

Another supporting evidence has also been brought by some recent studies on the environmental 
dependence of dust {\it extinction} levels of galaxies. \citet{2013MNRAS.434..423K} found a positive 
correlation ($\sim0.5$ mag level) between dust extinction ($A_{\rm H\alpha}$) and local density 
for $z=0.4$ SF galaxies, by comparing the IR- and H$\alpha$-based $SFR$s. They attributed the 
result to galaxy--galaxy interactions/mergers or gas/dust stripping resulting in a more 
compact configuration of star formation for SF galaxies residing in high-density environments.
This environmental dependence of dust extinction levels of SF galaxies has recently been 
confirmed with Balmer decrement analysis with optical spectroscopy \citep{2016MNRAS.tmp..323S}. 
We believe that these results may also be linked to the increment of $T_{\rm dust}$ with 
local density suggested by our current work.

However, the environmental impacts on dust properties of galaxies are still controversial. 
\citet{2015arXiv151100584N} performed FIR studies of $z\sim1.2$ clusters, and showed 
that the dust temperature does not strongly correlate with environment, except for a 
$\sim4\sigma$ drop in the average $T_{\rm dust}$ in an intermediate-density environment. 
They interpreted this result as invoking ram-pressure stripping of the warmer dust and reheating 
of the cold dust by the radiation of new stars which are formed by the surviving molecular gas.
However, we cannot confirm such a sharp decline of $T_{\rm dust}$ at any environment bin at least 
for our $z=0$ SF galaxies sample. 
\citet{2016MNRAS.tmp...95F} showed that the dust temperatures of late-type galaxies in the 
Coma cluster are hotter than those in the filamentary structure around the cluster, but the difference is small ($\lesssim1$ K) 
and has no statistical significance.
We should note, however, that our highest local density region 
(D5) covers relatively wide environmental range from clusters to surrounding filaments 
(see Fig.~\ref{RADEC}), 
and so a direct comparison between our results and those derived from studies 
on individual clusters would not be possible. 

Regarding the environmental effects on dust extinction, \citet{2011ApJ...735...53P} showed that 
the dust extinction ($A_V$) for SF galaxies from SED fitting declines by $\sim0.5$ mag 
from low to high local density at $0.6<z<0.9$.
On the other hand, \citet{2010MNRAS.402.2017G} showed that 
$A_{\rm H\alpha}$ has no significant dependence on environment for H$\alpha$-selected galaxies
at $z=0.84$. In these ways, environmental effects on dust properties of galaxies are 
still under debate. More studies are clearly needed to understand the environmental effects on dust properties of galaxies.

We finally comment that it is not possible to determine $\beta$ value with our current dataset. 
There is no study which explicitly reported environmental dependence (or independence) of $\beta$, and actually 
it is very hard to determine $T_{\rm dust}$ and $\beta$ separately because $T_{\rm dust}$ and $\beta$ are \textcolor{blue}{degenerate}.
As we briefly discussed in Section \ref{DerivationTd}, the estimated $T_{\rm dust}$ can be slightly increased by assuming smaller $\beta$ value.
Therefore the marginal trend between $T_{\rm dust}$ and $\rho_5$ could be invoked by a reduction of $\beta$ with increasing $\rho_5$.
More detailed studies on $\beta$ are needed to determine the cause of the relation between $T_{\rm dust}$ and $\rho_5$.

\section{Summary}\label{summary}
In this paper, we present infrared views of the environmental effects
on dust properties in star-forming (SF) galaxies at $z\sim0$. In order
to reveal the effects statistically, we use the AKARI FIS all-sky map
and the large spectroscopic galaxy sample from SDSS DR7. We define the
normalized local galaxy density ($\rho_5$) for each galaxy, which
represents the ratio of its fifth nearest neighbour surface density to
the mean density for all galaxies within a redshift slice of $\Delta z
= \pm 0.003$ in SDSS. In this study, we use galaxies within the
redshift range of $0.05<z<0.07$ and the stellar mass range of $9.2 <
\log_{10}(M_*/M_\odot)$. We select SF galaxies based on their
H$\alpha$ equivalent widths ($EW_{\rm H\alpha}>4$ \AA) and emission
line flux ratios using the BPT diagram. We then split them into five
local density bins: D1--D5.

In order to investigate average FIR properties of SF galaxies, we
perform FIR stacking analyses by splitting the SDSS SF galaxy sample
according to their stellar mass, $(S)SFR$, and environment. We derive
total infrared luminosity ($L_{\rm IR}$) for each subsample using the
average flux densities at WIDE-S (90 $\mu$m) and WIDE-L (140 $\mu$m)
bands, and then compute IR-based $SFR$ ($SFR_{\rm IR}$) from $L_{\rm
IR}$. We find that $SSFR_{\rm IR}$ of SF galaxies decreases
monotonically from the low to high density regions. We note that the
environmental difference is not large ($\sim0.1$ dex level even if we
compare the lowest- and highest-density bins), but this decline 
in star formation activity {\it amongst SF galaxies} suggests that
the environmental effects do not simply shut down the SF activity
instantly, implying a slow quenching mechanism (e.g.\ strangulation)
at work over wide environmental range.

We also derive average dust temperature ($T_{\rm dust}$) of SF
galaxies using the flux densities at 90 $\mu$m and 140 $\mu$m bands.
We study the dependence of $T_{\rm dust}$ on galaxy properties, 
and confirm a strong positive correlation between $T_{\rm dust}$ and 
$SSFR_{\rm IR}$, consistent with recent studies.
We investigate the environmental impacts on the average
$T_{\rm dust}$ of SF galaxies, and find an interesting hint that
$T_{\rm dust}$ increases with increasing environmental density.
Although the environmental trend is much milder (and only marginal)
than the $SSFR$--$T_{\rm dust}$ correlation, our results suggest that
the environmental effects may affect dust temperature in SF galaxies.
The physical mechanism which is responsible for this phenomenon is not
clear, but we suggest that it is not necessarily a cluster-specific
mechanism because $T_{\rm dust}$ monotonically increases from lowest-
to highest-density environments. We note that our results do not
change even if we consider the small environmental difference in $SSFR$;
i.e\ we confirm that the offset value from the $SSFR$--$T_{\rm dust}$
relation ($\Delta T_{\rm dust}$) shows the same environmental trend as
that for $T_{\rm dust}$.

This paper provides the first systematic study on the environmental
dependence of the dust temperature in SF galaxies by taking advantage
of the wide-field (all-sky) coverage at FIR with the newly released
AKARI FIS map. We find a marginal, but potentially an important
hint that the average $T_{\rm dust}$ of SF galaxies 
may increase with environmental density.
We note that the weak environmental trend could also be caused by a reduction of $\beta$ with increasing $\rho_5$, but in any case, our study suggests that dust properties in SF galaxies (dust temperature and/or dust composition) may depend on environment.
More detailed studies on individual galaxies, 
particularly spatially resolved studies of dust properties within 
the galaxies (with high-resolution FIR--submm observations), 
are needed to identify physical properties that produce 
the environmental trend we reported in this paper.

\section*{Acknowledgments}
We thank the referee for reviewing our paper and giving us valuable advice which improved the paper.
This research is based on observations with AKARI,
a JAXA project with the participation of ESA.
We thank Prof. Kotaro Kohno and Prof. Hideo Matsuhara for their valuable advice to our study.
This work was financially supported in part by a
Grant-in-Aid for the Scientific Research (No. 25247016, 26247030, 26800107)
by the Japanese Ministry of Education, Culture, Sports and Science.




\bibliographystyle{mnras}
\bibliography{references} 



\appendix


\bsp	
\label{lastpage}
\end{document}